\documentclass[a4paper,11pt]{article}

\usepackage{jheppub1} 

\usepackage[latin3]{inputenc}
\usepackage[makeroom]{cancel}
\usepackage{graphicx}
\usepackage{amsmath}
\usepackage{amsfonts}
\usepackage{amssymb}
\usepackage{color}
\usepackage[export]{adjustbox}
\usepackage{graphicx}
\usepackage{dcolumn}
\usepackage{bm}
\usepackage{simplewick}
\usepackage{array}
\usepackage{appendix}
\usepackage{relsize}
\usepackage{subcaption}

\newcommand{\be}{\begin{equation}}
    \newcommand{\ee}{\end{equation}}
\newcommand{\beq}{\begin{equation}}
    \newcommand{\eeq}{\end{equation}}
\newcommand{\bea}{\begin{eqnarray}}
    \newcommand{\eea}{\end{eqnarray}}

\title{\boldmath Charging the Johannsen-Psaltis spacetime}

\author{Rehana Rahim and}
\author{Khalid Saifullah}

\affiliation {Department of Mathematics, Quaid-i-Azam University, Islamabad, Pakistan}

\emailAdd{rehana.rahim8@gmail.com}
\emailAdd{ksaifullah@fas.harvard.edu}
\abstract{In this paper we develop the charged version of the spacetime proposed
by Johannsen and Psaltis. Rotation is introduced in the deformed Reissner-Nordstr\"om spacetime by complex
coordinate transformation. The event horizon and Killing horizon are studied. Killing horizons
 are represented graphically also. The analysis of the determinant of the metric shows that the spacetime does not
 have Lorentz violating regions. Similarly, from the study of the closed time-like curves, we see that no such curves exist outside the central
 body. Expressions for the energy and angular momentum for a
 particle on the equatorial plane are determined. Location of the circular photon orbits
 and innermost stable circular orbits is also computed.
\vspace{75 mm} }

\bigskip\bigskip

\notoc

\begin{document}
\maketitle
\flushbottom

\section{Introduction}

It is a difficult task to find rotating solutions in the general
theory of relativity which was
given by Einstein in 1915. The van Stockum rotating
solution \cite{vs} was developed two decades later
and it took more than forty years to discover the rotating solution of Kerr.
According to the no-hair theorem \cite{2a}, solutions to the Einstein-Maxwell
theory are uniquely described by the three parameters namely mass, charge and rotation.  In the
static and spherically symmetric setting the Schwarzschild and Reissner-Nordstr\"om are
solutions to the Einstein field equations. The Kerr metric is the rotating generalization of
the Schwarzschild solution and Kerr-Newman spacetime is the
rotating generalization of the Reissner-Nordstr\"om solution as well as charged generalization
of the Kerr metric.
The Newman-Janis algorithm \cite{nj} was used to develop exterior rotating
solutions but later was applied for developing rotating interior metrics which were matched to
the exterior Kerr \cite{intro1}. This algorithm is based on complex coordinate transformation. It
introduces rotation into the static spacetimes.  Kerr and Kerr-Newman spacetimes have been
derived via Newman-Janis algorithm by using the Schwarzschild and Reissner-Nordstr\"om spacetimes as seed metrics respectively \cite{nj,nj1}.

The general theory of relativity has been put to test extensively in numerous
ways. One such test is to see that the black holes are described by the Kerr solution.
This can be done in many ways \cite{2b}. Tests in the weak gravitational field regime can depend on the parameterized post-Newtonian
approach \cite{3}.
On the contrary, one must model the black
hole metric in terms of parametric deviations from the Kerr spacetime in order
to take the test in strong gravitational regime. A number of such spacetimes
have been developed \cite{4b,jp,chern}.

A particularly interesting modified Kerr spacetime has been proposed by Johannsen
and Psaltis \cite{jp} by applying the Newman-Janis algorithm to a deformed Schwarzschild metric. It is a
stationary, axisymmetric, asymptotically flat vacuum solution of some
unknown field equations differing from the Einstein field equations due to the presence
of function $h(r,\theta)$. In
addition to the spin and mass parameters, this metric contains at least one parameter which
measures the potential deviations from the Kerr spacetime. Setting such deviation parameters to
zero, one recovers the Kerr metric. It could be
the ad-hoc starting point for the investigation
of deviations from general relativity \cite{han}. In case of modified theories of gravity, Newman-Janis formulation
usually does not relate static solutions with corresponding stationary solutions but can be applied to modified
forms of the Schwarzschild spacetime \cite{yunes}.

Most known astrophysical compact objects are either electrically
neutral or slightly charged. The Kerr-Newman black hole is an exact
electro vacuum solution of the Einstein-Maxwell system of equations.
It is an ideal model for the interaction between the electromagnetic
field and the gravitoelectric and gravitomagnetic components of
gravity. Therefore, it is significant both conceptually and
theoretically. Keeping in view the Kerr-Newman metric of general relativity, we look
for a rotating charged spacetime in the alternate theories of gravity. In the
case of vector-tensor theories of alternate gravity, an exact
rotating solution has been developed recently \cite{ft}. The charged
rotating solutions in the limit of small Weyl corrections was
developed in Ref. \cite{weyl}.

The main motivation for this work comes from the Johannsen-Psaltis
spacetime. In this paper we take the deformed Reissner-Nordstr\"om
spacetime as the seed metric and construct the charged
generalization of the Johannsen-Psaltis spacetime.

This paper is organized as follows. In Section \ref{TD} the charged
metric is developed. In Section \ref{H} the event horizon and
Killing horizon have been discussed. Section \ref{LCR} analyzes
our metric for the existence of Lorentz violation regions and closed
time-like curves. The innermost stable circular
orbits (ISCO) and circular photon orbits are computed in Section \ref{PM}. The work is
concluded in the last section.

\section{The construction}\label{TD}
The Reissner-Nordstr\"om metric is given by
\begin{equation}
ds^{2}=-\left(1-\frac{2M}{r}+\frac{q^2}{r^2}\right) dt^{2}+\left(1-\frac{2M}{r}+\frac{q^2}{r^2}\right)^{-1}dr^{2}+r^2 d\theta^{2}+r^2 \sin^{2}\theta d\phi^{2},
\label{1}
\end{equation}
where $M$ is the mass of the central object having charge $q$. The 4-potential for the above metric is
\begin{equation}
A_\mu=\left(-\frac{q}{r},0,0,0\right).
\label{1a}
\end{equation}
The $(t-r)$ sector is modified by multiplying the corresponding component by the expression of the form $1+h(r)$ where $h(r)$ is given by \cite{jp}
\begin{equation}
h(r)=\sum_{k=0}^{\infty}\epsilon_{k}\left(\frac{M}{r}\right)^k.
\label{2}
\end{equation}
The deformed Reissner-Nordstr\"om spacetime thus takes the form
\begin{align}
ds^{2}&=-\left(1-\frac{2M}{r}+\frac{q^2}{r^2}\right)\Big(1+h(r)\Big) dt^{2}+\left(1-\frac{2M}{r}+\frac{q^2}{r^2}\right)^{-1}\Big(1+h(r)\Big) dr^{2}+r^2 d\theta^{2}
\nonumber
\\
&
+r^2 \sin^{2}\theta d\phi^{2}.
\label{3}
\end{align}
Setting $\epsilon_{k}=0$ gives the Reissner-Nordstr\"om spacetime. First, let us denote $f=1-2M/r+q^2/
r^2$ for simplification. The next step is to change from $(t,r,\theta,\phi)$ to the Eddington-
Finkelstein coordinates  $(u^\prime,r^\prime,\theta^\prime,\phi^\prime)$ where
\begin{align}
&du^{\prime}=dt-\frac{dr}{f},
\label{4}
\\
&
r=r^\prime,\quad
\theta=\theta^\prime,\quad
\phi=\phi^\prime.
\label{5}
\end{align}
Using the above equations, Eq. (\ref{3}) takes the form
\begin{equation}
g_{\mu\nu}=-f\Big(1+h(r)\Big) du^{2}-2\Big(1+h(r)\Big) dudr+r^2 d\theta^{2}+r^2 \sin^{2}\theta d\phi^{2},
\label{6}
\end{equation}
having inverse
\begin{equation}
g^{\mu\nu}=-\frac{2}{\Big(1+h(r)\Big)}dudr+\frac{f}{\Big(1+h(r)\Big)}dr^2+\frac{1}{r^2} d\theta^{2}+\frac{1}{r^2 \sin^{2}\theta} d\phi^{2},
\label{7}
\end{equation}
where we have removed the primes.
The above inverse metric can also be expressed in
the Newman-Penrose formalism \cite{np} as
\begin{equation}
g^{\mu\nu}=-l^{\mu}n^{\nu}-l^{\nu}n^{\mu}+m^{\mu}\overline{m}^{\nu}+m^{\nu}\overline{m}^{\mu},
\label{8}
\end{equation}
using the null vectors. For the metric in Eq. (\ref{7}) these are given as
\begin{align}
&l^{\mu}=\Big(0,1,0,0\Big), \quad n^{\mu}=\frac{1}{\Big(1+h(r)\Big)}\Big(1,\frac{-f}{2},0,0\Big),
\label{9}
\\
&
m^{\mu}=\frac{1}{\sqrt2 r}\Big(0,0,1,\frac{i}{\sin\theta}\Big), \quad \overline{m}^{\mu}=\frac{1}{\sqrt2 r}\Big(0,0,1,-\frac{i}{\sin\theta}\Big).
\label{10}
\end{align}
The indices have range $0-3$. Next, we consider $r$ to be complex and re-write the above null vectors as
\begin{align}
&l^{\mu}=\Big(0,1,0,0\Big), \quad  n^{\mu}=\frac{1}{\Big(1+h(r,\overline{r})\Big)}\Big(1,\frac{-f(r,\overline{r})}{2},0,0\Big),\label{13}
\\
&
m^{\mu}=\frac{1}{\sqrt2 r}\Big(0,0,1,\frac{i}{\sin\theta}\Big), \quad   \overline{m}^{\mu}=\frac{1}{\sqrt2 \overline{r}}\Bigg(0,0,1,-\frac{i}{\sin\theta}\Bigg),\label{15}
\end{align}
where bar denotes the complex conjugate and the expressions for $f(r,\overline{r})$ and $h(r,\overline{r})$ are
\begin{align}
&f(r,\overline{r})=1-M\left(\frac{1}{r}+\frac{1}{\overline{r}}\right)+\frac{q^2}{r\overline{r}},
\\
&
 h(r,\overline{r})=\sum_{k=0}^{\infty}\left(\epsilon_{2k}+
\epsilon_{2k+1}\frac{M}{2}\left(\frac{1}{r}+\frac{1}{\overline{r}}\right)\right)\left(\frac{M^2}{r\overline{r}}\right)^{k}. \label{17}
\end{align}
Using the transformation
\begin{align}
&u^\prime=u-i a \cos\theta, \quad r^\prime=r+i a \cos\theta,
\label{18}
\\
&
\theta=\theta^\prime,\quad \phi=\phi^\prime,
\label{19}
\end{align}
the null vectors in Eqs. (\ref{13})-(\ref{17}) take the form
\begin{align}
&l^{\mu}=\Big(0,1,0,0\Big), \quad n^{\mu}=\frac{1}{\Big(1+h(r^\prime,\theta^\prime)\Big)}\Big(1,\frac{-f(r^\prime,\theta^\prime)}{2},0,0\Big),\label{21}
\\
&
m^{\mu}=\frac{1}{\sqrt2 r^\prime}\Big(i a\sin\theta^\prime,-i a\sin\theta^\prime,1,\frac{i}{\sin\theta^\prime}\Big),\label{22}
\\
&
\overline{m}^{\mu}=\frac{1}{\sqrt2 \bar{r}^\prime}\Big(-i a\sin\theta^\prime,i a\sin\theta^\prime,1,-\frac{i}{\sin\theta^\prime}\Big).\label{23}
\end{align}
Here the expressions for $f(r^\prime,\theta^\prime), h(r^\prime,\theta^\prime)$ are
\begin{align}
&f(r^\prime,\theta^\prime)=1-\frac{2Mr^\prime}{\Sigma}+\frac{q^2}{\Sigma},\label{25}
\\
&
h(r^\prime,\theta^\prime)=\sum_{k=0}^{\infty}\left(\epsilon_{2k}+
\epsilon_{2k+1}\frac{Mr^\prime}{\Sigma}\right)\left(\frac{M^2}{\Sigma}\right)^{k}, \label{26}
\end{align}
where $\Sigma$ is given by 
\begin{equation}
\Sigma=r^{\prime2}+a^2\cos^{2}\theta^\prime.
\label{24}
\end{equation}
Using Eqs. (\ref{21})-(\ref{24}), the metric tensor $g_{\mu\nu}$ is written in terms of
 the coordinates $(u^\prime,r^\prime,\theta^\prime,\phi^\prime)$ as
\begin{align}
&g_{00}=-f\left(1+h\right),\quad g_{01}=-\left(1+h\right),\label{27}
\\
&
g_{03}=a\left(1+h\right)\left(f-1\right)\sin^{2}\theta^\prime,\quad g_{13}=a\left(1+h\right)\sin^{2}\theta^\prime,\label{28}
\\
&
g_{22}=\Sigma,\quad  g_{33}=\sin^{2}\theta^\prime\left[\Sigma-a^2\left(1+h\right)\left(f-2\right)\sin^{2}\theta^\prime\right].\label{29}
\end{align}
In order to remove the off-diagonal terms $g_{01}$ and $g_{13}$, we use the transformation \cite{jp,cav}
\begin{align}
&du^\prime=dt-W(r^\prime,\theta^\prime)dr,\quad d\phi^\prime=d\phi-G(r^\prime,\theta^\prime)dr,\label{30}
\\
&
r=r^\prime,\quad \theta=\theta^\prime,\label{31}
\end{align}
where
\begin{equation}
W(r^\prime,\theta^\prime)=\frac{g_{01}g_{33}-g_{03}g_{13}}{g_{00}g_{33}-g_{03}^2}, \quad G(r^\prime,\theta^\prime)=\frac{g_{00}g_{13}-g_{01}g_{03}}{g_{00}g_{33}-g_{03}^2}.\label{32}
\end{equation}
This leads to the metric tensor given by
\begin{eqnarray}
ds^{2} &=&-(1+h(r,\theta ))\Big(1-\frac{2Mr}{\Sigma }+\frac{q^2}{\Sigma }\Big)dt^{2}-\frac{2a(2Mr-q^2)\sin
^{2}\theta }{\Sigma }(1+h(r,\theta ))dtd\phi
\notag \\
&&
+\frac{\Sigma (1+h(r,\theta ))}{
\Delta +a^{2}\sin ^{2}\theta h(r,\theta )}dr^{2} +\Sigma d\theta ^{2}+\Big[\sin ^{2}\theta \Big(r^{2}+a^{2}+\frac{a^2(2Mr-q^2)\sin
^{2}\theta }{\Sigma }\Big)
\notag \\
&&
+h(r,\theta )\frac{a^{2}\sin ^{4}\theta (\Sigma +2Mr-q^2)}{
\Sigma }\Big]d\phi ^{2},  \label{33}
\end{eqnarray}
where $\Sigma =r^{2}+a^{2}\cos ^{2}\theta ,$ $\Delta =r^{2}+a^{2}-2Mr+q^2$ and $%
h(r,\theta )$ has the general expression given in Eq. (\ref{26}) with $r=r^\prime$ and $\theta=\theta^\prime$. Setting
 $q=0$ gives the Johannsen-Psaltis metric. With $h=0$, the Kerr-Newman spacetime is recovered.

The function $h(r,\theta )$ contains infinite number of parameters. The first two parameters
$\epsilon_0$ and $\epsilon_1$ are set to zero by requiring that the metric must be asymptotically
flat and the next parameter $\epsilon_2$ is constrained at $10^{-4}$ by weak field tests of general
relativity in the parameterized post-Newtonian approach. Thus $\epsilon_2$ can also be set to zero. As in the case of Ref. \cite{jp}, we set  $\epsilon_k=0$ for $k>3$, which leads to $h(r,\theta )$  as
\begin{equation}
h(r,\theta )=\frac{\epsilon_{3}M^{3}r}{\Sigma^2},\label{h}
\end{equation}
which is the same as in Ref. \cite{jp}. Since $\epsilon_3$ is the only retained parameter, we drop the subscript 3 in the rest of the paper and represent it as $\epsilon$.
Applying the Newman-Janis algorithm on the potential $A_\mu$ in Eq. (\ref{1a}) leads to
\begin{equation}
A_\mu=\left(-\frac{qr}{\Sigma},\frac{qr}{\Delta+a^2h\sin ^{2}\theta},0,\frac{aqr\sin ^{2}\theta}{\Sigma}\right).\label{pot1}
\end{equation}
Here the $A_r$ component depends on $\theta$. So it cannot be gauged away. This makes the electromagnetic
potential in Eq. (\ref{pot1}) different from  Kerr-Newman's potential.

Expanding the $A_r$ component in terms of powers of the deviation
parameter $\epsilon$, we obtain
\begin{equation}
A_r=\frac{qr}{\Delta} \big(1-\frac{a^2\epsilon M^{3}r\sin ^{2}\theta
}{\Sigma^2}+ \text{higher order terms}\big).\label{pot2}
\end{equation}
From the second term we see that the electromagnetic potential
depends on mass $M.$  Thus we can consider the deviation parameter
as a coupling between the gravitational and electromagnetic fields.

Thus we see that the Maxwell tensor has the components
\begin{align}
&F_{tr}=-\frac{q \left(r^2-a^2 \cos ^2\theta
\right)}{\Sigma^2},\quad F_{t\theta}=\frac{a^2 q r \sin2 \theta
}{\Sigma^2},\label{max1}
 \\ & F_{r\theta}=-\frac{a^2 M^3 q r^2 \epsilon  \sin
\theta \cos \theta  \left(a^2 \cos 2 \theta -3 a^2-2
r^2\right)}{\Sigma^3 (\Delta +a^2 h \sin ^2\theta)^2},\label{max2}
\\ &
F_{\phi r}=-\frac{a q \sin ^2\theta \left(a^2 \cos ^2\theta
-r^2\right)}{\Sigma^2},\quad F_{\phi \theta}=-\frac{a q r
\left(a^2+r^2\right) \sin 2 \theta}{\Sigma^2}. \label{max3}
\end{align}

From these expressions of the Maxwell tensor we note that the only
component different from the Kerr-Newman's is the $F_{r\theta}$
component. The metric tensor in Eq. (\ref{33}) with $h(r,\theta)$ as
in Eq. (\ref{h}) and the Maxwell tensor components (Eqs.
(\ref{max1})-(\ref{max3})) do not satisfy the usual Einstein-Maxwell
equations. So, we assume that our metric is an electro vacuum
solution to some unknown field equations which are different from
the Einstein-Maxwell equations for nonzero $h(r,\theta)$.
\section{The horizons}\label{H}
The event horizon of the spacetime (\ref{33}) (if it exists) is located at a radius $r_{hor}=H(\theta)$, where $H(\theta)$ is obtained by solving the first order ordinary differential equation \cite{sys}
\begin{equation}
g^{rr}-2g^{r\theta}\left(\frac{dH}{d\theta}\right)+g^{\theta\theta}\left(\frac{dH}{d\theta}\right)^2=0,\label{34a}
\end{equation}
where $g^{rr}=1/g_{rr}$ and $g^{\theta\theta}=1/g_{\theta\theta}$. As $g^{r\theta}=0$ for the spacetime under consideration, so Eq. (\ref{34a}) reduces to
\begin{equation}
g^{rr}+g^{\theta\theta}\left(\frac{dH}{d\theta}\right)^2=0. \label{34}
\end{equation}
For the case of $\theta=0, \pi$ and equatorial plane $\theta=\pi/2$, Eq. (\ref{34}) reduces to $g^{rr}=0$.
For the pole $\theta=0$ or $\theta=\pi$, $g^{rr}\propto\Delta$
having root $r=H_{KN}=r_{+}$ where $H_{KN}$ is the event horizon of the Kerr-Newman spacetime. This is possible if $\epsilon\neq-(2Mr_{+}-q^2)^2/M^3r_{+}$ at which the denominator of $g^{rr}$ vanishes.
  Here we will expand the function $H(\theta)$ \cite{sys}, first around $\theta=0$ and then the same analysis will be done for  $\theta=\pi/2.$
Let  $\theta=0+\delta\theta$. In this case $r=H(\delta\theta)$ is given by
\begin{equation}
H(\delta\theta)=H_{KN}+\delta\theta\frac{dH}{d\theta}\mid_{\theta=0}+\frac{\delta\theta^2}{2}\frac{d^2H}{d\theta^2}\mid_{\theta=0}.\label{35}
\end{equation}
Putting $r=H(\theta)$ in Eq. (\ref{34}) and expanding in terms of $\theta=\delta\theta$, we obtain for the first order in $\delta\theta$,
\begin{align}
&\frac{\left(\frac{dH}{d\theta}\mid_{\theta=0}\right)^2 }{\left(2MH_{KN}-q^2\right)}
+\delta \theta\Bigg[-2 \left(\frac{dH}{d\theta}\mid_{\theta=0}\right)^3 H_{KN} 
\nonumber
\\
&
+2\frac{dH}{d\theta}\mid_{\theta=0} \Big(2MH_{KN}-q^2
\Big)\Big(\frac{\sqrt{-a^2+M^2-q^2}}{1+\frac{\epsilon M^3 H_{KN}}{(2MH_{KN}-q^2)^2}
}+\frac{d^2 H}{d\theta
^2}\mid_{\theta=0}\Big)\Bigg]=0\label{36}.
\end{align}
From the above equation it can be concluded that
\begin{equation}
\frac{dH}{d\theta}\mid_{\theta=0}=0.
\end{equation}
For the second order, $\delta\theta^2$, we obtain the equation for $\frac{d^2 H}{d\theta
^2}\mid_{\theta=0}$ as
\begin{align}
(\frac{d^2H}{d\theta^2}\mid_{\theta=0})^2 \Big(1+\frac{\epsilon M^3 H_{KN}}{(2MH_{KN}-q^2)^2}\Big)+\frac{a^2\epsilon M^3 H_{KN}}{(2MH_{KN}-q^2)^2}
+\sqrt{-a^2+M^2-q^2} \frac{d^2H}{d\theta^2}\mid_{\theta=0}=0.\label{37}
\end{align}
For the existence of null surface it is required that $\frac{d^2 H}{d\theta
^2}\mid_{\theta=0}$ must have a real solution. This requirement leads to an upper and lower bound for the deviation parameter $\epsilon$ as
\begin{align}
&\epsilon^{max-pole}=-\frac{\beta}{2M^3H_{KN}}+\frac{\beta\sqrt{M^2-q^2}}{2M^3aH_{KN}},\label{38}
\\
&
\epsilon^{min-pole}=-\frac{\beta}{2M^3H_{KN}}-\frac{\beta\sqrt{M^2-q^2}}{2M^3aH_{KN}},\label{39}
\end{align}
where $\beta=2MH_{KN}-q^2$.

On the equatorial plane $\theta=\pi/2$, the equation $g^{rr}=0$, takes the form
\begin{equation}
r^2+a^2-2Mr+q^2+\frac{a^2\epsilon M^3}{r^3}=0.\label{40}
\end{equation}
At the equatorial plane, the maximum positive value of the parameter $\epsilon$ for the real solution of Eq. (\ref{40}) is given by
\begin{align}
\epsilon^{max-eq}&=\frac{1}{3125a^2 M^2}\Bigg[150 (\alpha +15) a^4-20 a^2 [8(7 \alpha +40) M^2-15 (\alpha +15) q^2]
\nonumber
\\
&
+1024 (\alpha +4) M^4
-160 (7 \alpha+40)M^2 q^2+150(\alpha +15)q^4\Bigg],\label{41}
\end{align}
where $\alpha=\sqrt{16-\frac{15 a^2}{M^2}-\frac{15 q^2}{M^2}}.$ Thus the null surface at the equatorial plane exists for the deviation parameter in the range
$\epsilon<\epsilon^{max-eq}$. By comparing maximum values of $\epsilon$ at the pole and the equatorial plane we find that $\epsilon^{max-pole}<\epsilon^{max-eq}$.

For the case of the equatorial plane, it is assumed that $\theta=\pi/2+\delta\theta$
so that $H(\theta)$ is given by
\begin{equation}
H(\pi/2+\delta\theta)=H_{eq}+\delta\theta\frac{dH}{d\theta}\mid_{\theta=\pi/2}+\frac{\delta\theta^2}{2}\frac{d^2H}{d\theta^2}\mid_{\theta=\pi/2},\label{42}
\end{equation}
where $H_{eq}$ is the radius of the null surface at ${\theta=\pi/2}$. Substituting $r=H(\pi/2+\delta\theta)$ in Eq. (\ref{34}), and
expanding in terms of $\delta\theta$, we find that up to first order in  $\delta\theta$, $\frac{dH}{d\theta}\mid_{\theta=\pi/2}=0.$ In
the second order in $\delta\theta$, the real solution for $\frac{d^2H}{d\theta^2}\mid_{\theta=\pi/2}$ exists
 for all $\epsilon$ in the range $0<\epsilon<\epsilon^{max-pole}$
but same cannot be said for the range $\epsilon^{min-pole}<\epsilon<0$.

\subsection{Killing horizon}\label{KH}
A Killing horizon is a null hypersurface on which there is a
null Killing vector field. In a stationary and axisymmetric
spacetime, the Killing horizon is given by
\begin{equation}
g_{tt}g_{\phi\phi}-g^{2}_{t\phi}=0,\label{kill}
\end{equation}
which for metric (\ref{33}) becomes  $(1+h)(\Delta+a^{2}h\sin^{2}\theta)\sin^{2}\theta=0$. For $\theta=0$ or $\theta=\pi$, we have
\begin{equation}
g_{tt}g_{\phi\phi}-g^{2}_{t\phi}\propto\left(r^2+a^2-2Mr+q^2\right)\left[1+\frac{\epsilon M^{3}r}{(r^2+a^2)^2}\right].\label{k1}
\end{equation}
As in the case of event horizon at the pole, the Killing horizon at
the pole coincides with the Kerr-Newman event horizon. From Eq.
(\ref{k1}) the value of $\epsilon$ comes out to be
\begin{equation}
\epsilon^{kil-pol}=-\frac{16}{3\sqrt{3}}\Big(\frac{a}{M}\Big)^3.
 \label{kA}
\end{equation}

For the case of the equatorial plane, Eq. (\ref{kill}) takes the form
\begin{equation}
(1+\frac{\epsilon M^{3}}{r^3})(r^2+a^2-2Mr+q^2+\frac{a^{2}\epsilon M^{3}}{r^3})=0.\label{k2}
\end{equation}
The second factor in the above equation is same as in Eq. (\ref{40}) having two solutions for $\epsilon<\epsilon^{max-eq}$ and single
solution for negative $\epsilon$. In the extremal case when $a^2+q^2>M$, Eq. (\ref{k2}) has solutions
for negative deviation parameter only. At the pole Eq. (\ref{k1})
has two solutions for $\epsilon<\epsilon^{kil-pol}$, one real solution
for $\epsilon=\epsilon^{kil-pol}$ and no solution otherwise.

The Killing horizon for different values of $\epsilon,a$ and $q$ are
shown in Figure 1.
\begin{figure}[!hptb]
\centering
\includegraphics[scale=0.8]{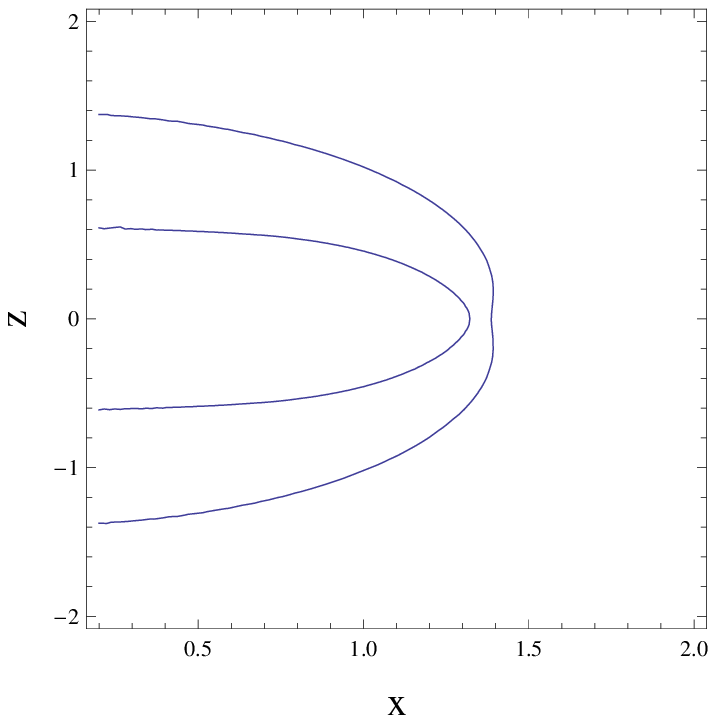}
\includegraphics[scale=0.8]{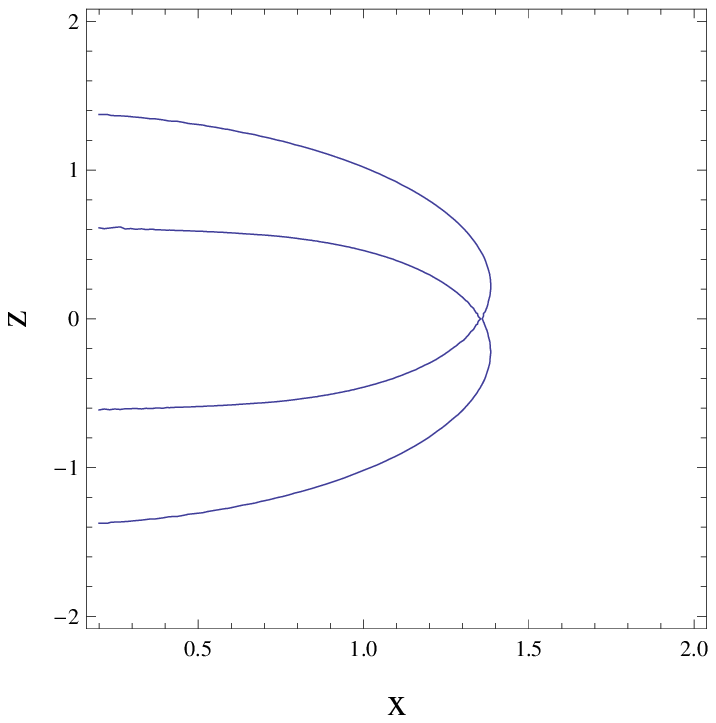}
\includegraphics[scale=0.8]{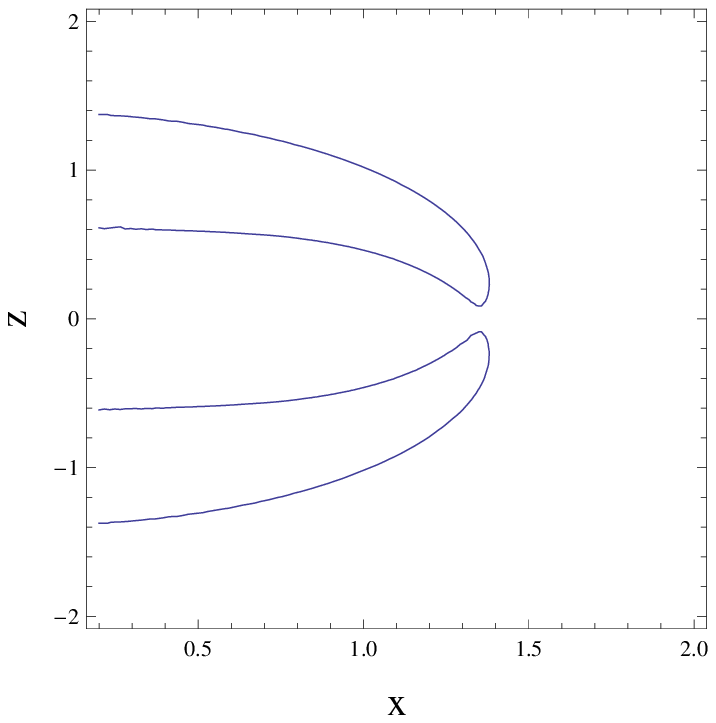}
\caption{Shapes of the Killing horizon for $a=0.7,q=0.6$ and $M=1$ with varying values of $\epsilon$ as $0.389, 0.396$ and $0.4$ (clockwise).}
\end{figure} 

On the top left $\epsilon=0.389<\epsilon^{max-eq}$, the inner and
outer Killing horizons have spherical topology. On the top right $\epsilon=0.396\approx\epsilon^{max-eq}$, the inner
and outer Killing horizons merge at the equatorial plane. In the bottom graph $\epsilon=0.4>\epsilon^{max-eq}$, disjoint
Killing horizons appear above and below the equatorial plane. Figure 2 represents the Killing horizons for negative $\epsilon$.
\begin{figure}[!ht]
\centering
\includegraphics[scale=0.8]{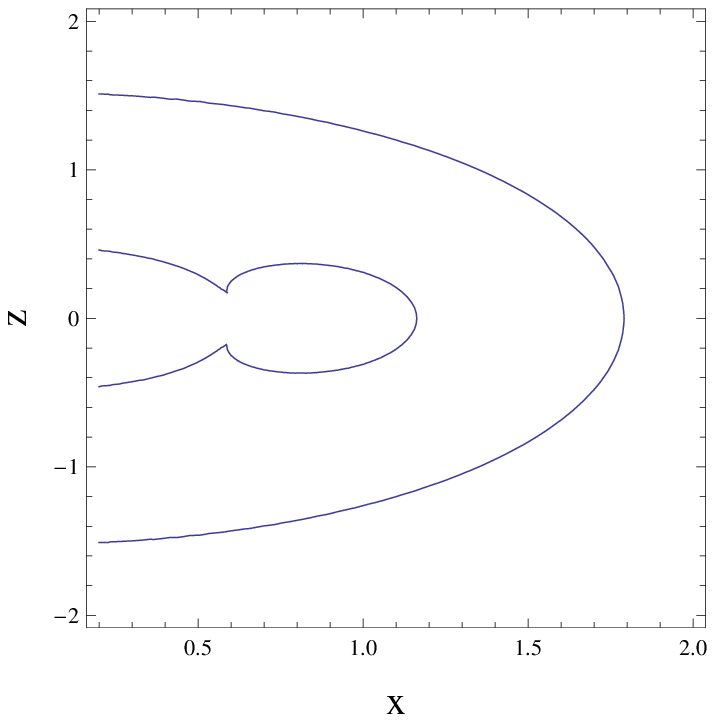}
\includegraphics[scale=0.8]{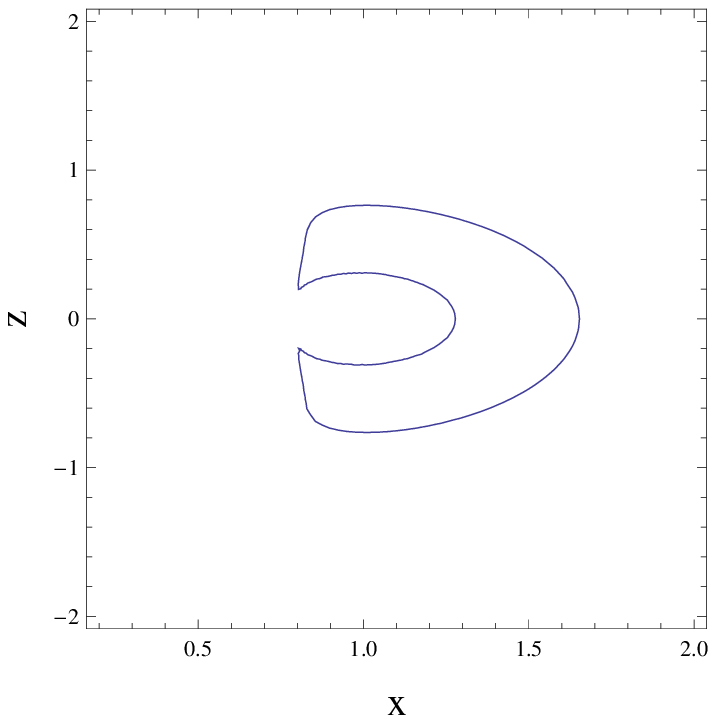}
\caption{The Killing horizon for negative $\epsilon$. On the left $a=0.8,q=0.3,\epsilon=-0.6$ and $M=1$.
On the right $a=0.96,q=0.43,\epsilon=-0.6$ and $M=1$.}
\end{figure}
In the left panel of the figure the values of the spin and charge
are chosen such that $a^2+q^2<M$. Here the inner and outer Killing horizons
 are in the form of spherical surface. In the right panel $a^2+q^2>M$. The Killing horizon is in the shape of toroidal surface in such case.

\subsection{Event horizon of the linearized charged Johannsen-Psaltis spacetime}\label{the approximate horizon}

The charged Johannsen-Psaltis metric can be treated as a small perturbation of the Kerr-Newman spacetime in terms of the deviation parameter $\epsilon$. The modifications in the Kerr-Newman metric components up to the first order in $\epsilon$ are 

 \begin{align}
&h_{tt}=-\frac{M^3 r  \left(\Sigma-2 M r+q^2 \right)}{\Sigma ^3}, \quad h_{rr}=\frac{M^3 r  \left(\Sigma-2 M r+q^2\right)}{\Sigma \Delta^2},
\nonumber
\\ 
&
h_{\theta\theta}=0,\quad h_{t\phi}=\frac{a M^3 r \sin ^2 \theta \left(q^2-2 M r\right)}{\Sigma ^3}, 
h_{\phi\phi}=\frac{a^2 M^3 r \epsilon  \sin ^4 \theta  \left(2 M r-q^2+\Sigma \right)}{\Sigma ^3}.
\end{align} 
The event horizon in this case can be determined from the equation \cite{sys}
\begin{equation}
g_{KN}^{rr}
\left(1-\epsilon g_{KN}^{rr}h_{rr}\right)=0,\label{43}
\end{equation}
where $g_{KN}^{rr}$ is the inverse of $g_{rr}$ component of the Kerr-Newman metric and $h_{rr}$ is given in the last equation.
The radius of the event horizon in this case is given by
\begin{equation}
r_H=H_{KN}\left(1+\lambda\epsilon\right),\label{44}
\end{equation}
where $H_{KN}$ is the event horizon of the Kerr-Newman black hole and $\lambda$ measures deviation from it.
Substituting Eq. (\ref{44}) in Eq. (\ref{43}), and linearizing in terms of $\epsilon$, we obtain
\begin{equation}
\lambda=-\frac{a^2M^3\sin^{2}\theta}{2\sqrt{M^2-a^2-q^2}(2MH_{KN}-q^2-a^2\sin^{2}\theta)^2}.\label{45}
\end{equation}
Thus the event horizon is located at the radius
\begin{equation}
r_H=H_{KN}\left(1-\frac{\epsilon a^2M^3\sin^{2}\theta}{2\sqrt{M^2-a^2-q^2}(2MH_{KN}-q^2-a^2\sin^{2}\theta)^2}\right).\label{46}
\end{equation}
The Kretschmann scalar for the linearized form of (\ref{33}) diverges at $r=0$, therefore, it represents a black hole.

\section{Lorentz violations, closed time-like curves and regions of validity}\label{LCR}

The determinant of metric (\ref{33}) is given by
\begin{equation}
det(g_{\mu\nu})=-\frac{\sin^{2}\theta}{64\Sigma^2} \Bigg[3a^4+8a^2r^2+8r^4+8\epsilon M^3r+4a^2(2r^2+a^2)\cos2\theta+a^4\cos4\theta\Bigg]^2,
\end{equation}
which is independent of the charge $q$. It is negative, semi-definite and becomes zero at two values of the radii for $\epsilon<-(2Mr_{+}-q^2)^2/M^3r_{+}$ which coincide with the location of the Killing horizon. So, the charged spacetime does not contain any Lorentz violating region.
\\
From the $g_{\phi\phi}$ component in metric (\ref{33}), we plot the closed time-like curves and combine them with the Killing horizons to show their location. These plots are shown in Figure 3. Also
from $g_{\phi\phi}$ component, we can determine the upper bound for $\epsilon$ at the equatorial plane and denote it with symbol $\epsilon^{CTC}$ and is given by
\begin{equation}
\epsilon^{CTC}=-\frac{r^3\left(r^2\left(a^2+r^2\right)-a^2\left(q^2-2 M r\right)\right)}{a^2 M^3 \left(r (2 M+r)-q^2\right)},
\end{equation}
while at the equatorial plane (\ref{40}) gives the value of $\epsilon$ as
 \begin{equation}
\epsilon^{hor}=-\frac{r^3\left(r^2+a^2+q^2-2Mr\right)}{a^2 M^3},
\end{equation}
where symbol $\epsilon^{hor}$ shows that this is the value at the horizon for the equatorial plane.
It is clear that $\epsilon^{CTC}<\epsilon^{hor}$ which leads to the result that for the equatorial plane the regions of closed time-like curves lie inside the inner Killing horizon.
\begin{figure}[!ht]
\centering
\includegraphics[scale=0.6]{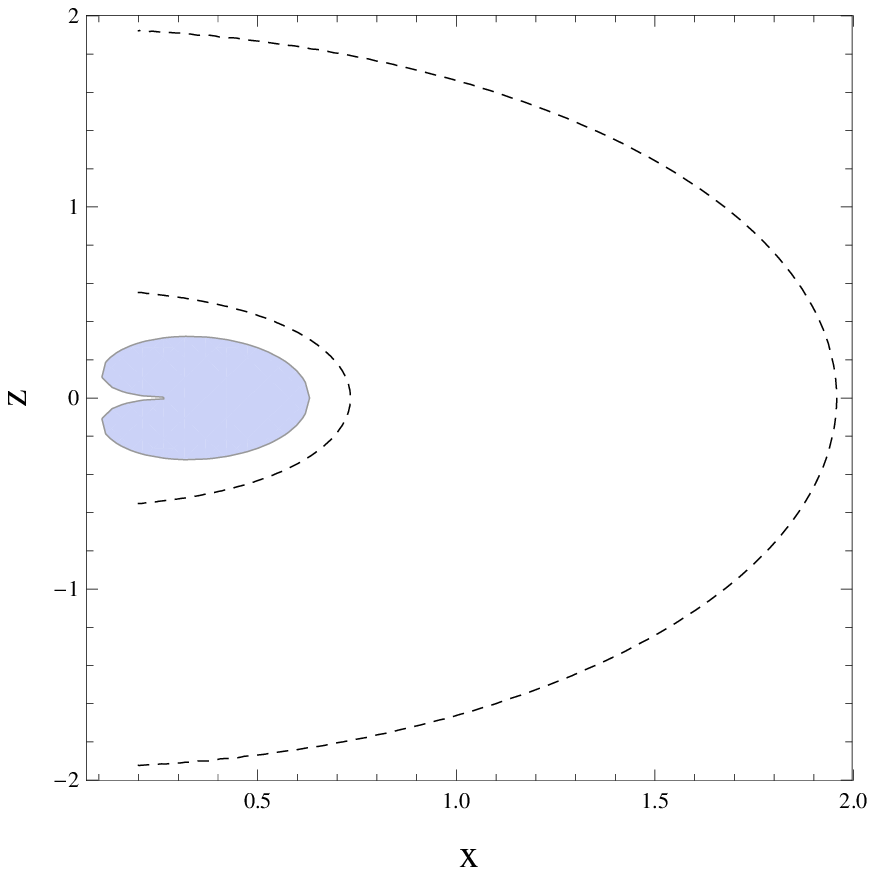}
\includegraphics[scale=0.6]{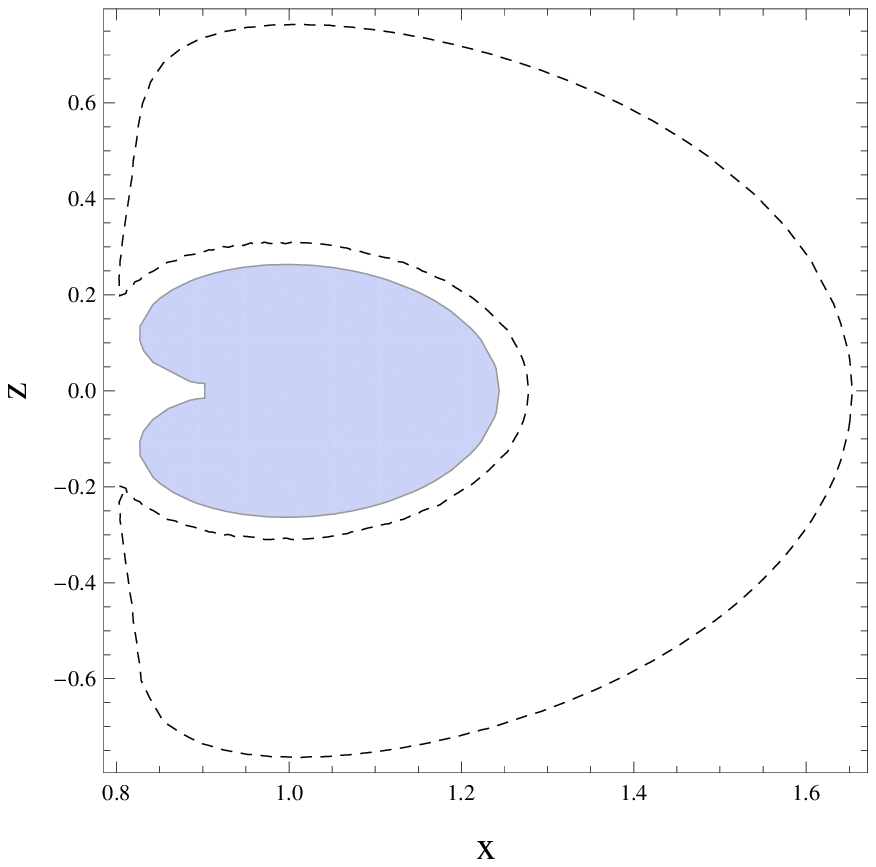}
\caption{The Killing horizon and closed time-like curves for some values of $\epsilon, a$ and $q$. Mass $M$ has been set to 1.  On the left $a=0.3,q=0.2,\epsilon=-0.3$.
On the right $a=0.96,q=0.43,\epsilon=-0.6$. The dashed curves denote the inner and outer Killing horizons and the solid region shows the closed time-like curve.}
\end{figure}
In Figure 3, two cases have been plotted. In the left panel the values of $q$ and $a$ are such that $a^2+q^2<M$, resulting in an inner and outer Killing horizon and the closed time-like region lying inside the inner Killing horizon. In the right panel $a^2+q^2>M$, resulting in the toroidal shaped horizon surrounding the closed time-like region.
So we conclude that the charged Johannsen-Psaltis metric does not contain closed time-like curves outside the central object.

\section{Innermost stable circular orbits and the circular photon orbits}\label{PM}

Rotation of the central object has significant effects on the motion
of particles in general relativity. Consider a massive particle
moving around a black hole. There exists a minimum radius at which
the stable circular motion is possible. This defines the innermost
stable circular orbits or the ISCO.

In this section, particle motion on the equatorial plane is
considered. First the expressions for the energy and angular
momentum are determined which are later employed to compute the
innermost stable circular orbits and circular photon orbits.

In order to obtain the radial equation of motion, we solve the
equation
\begin{equation}
p_{\alpha}p^{\alpha}=-m^2,
\end{equation}
for the radial momentum. Here $p_{\alpha}$ is the 4-momentum of the
particle of mass $m$. The radial equation of motion, so obtained is
\begin{equation}
(\frac{dr}{d\tau})^{2}=R(r)=-\frac{1}{g_{rr}}(g^{tt}E^2+g^{\phi\phi}L^2-2ELg^{t\phi}+m^2),
\end{equation}
where $\tau$ is the proper time. Here $E$ and $L$ denote the energy
and angular momentum of the particle, respectively, which can be
determined by solving the equations
\begin{align}
&R(r)=0, \\& \frac{dR(r)}{dr}=0.
\end{align}
This gives
\begin{equation}
\frac{E}{m}=\sqrt{\frac{P_{1}+P_{2}}{P_{3}}},
\end{equation}
where
\begin{align}
P_{1}&=12 a^4 M^3 r^5 \epsilon  \Big(M r-q^2\Big) \Big(M^3 \epsilon
+r^3\Big)^4+a^2 r^8 \Big(M^3 \epsilon +r^3\Big)^2 \Big[q^4 \Big(-7
M^6 \epsilon ^2+10 M^3 r^3 \epsilon \nonumber
\\
& +8 r^6\Big) +2 q^2 r \Big(16 M^7 \epsilon ^2-9 M^6 r \epsilon
^2-16 M^4 r^3 \epsilon +6 M^3 r^4 \epsilon -14 M r^6+6 r^7\Big)
\nonumber
\\
& +M r^2 \Big(-40 M^7 \epsilon ^2+48 M^6 r \epsilon ^2  -15 M^5 r^2
\epsilon ^2+16 M^4 r^3 \epsilon -6 M^2 r^5 \epsilon +20 M r^6-12
r^7\Big)\Big],
 \end{align}
\begin{align}
P_{2}&=2 \Big[2 \Big(r^{26}\mp P_{4}\Big)-48 M^{12} r^{14} \epsilon
^3+68 M^{11} r^{15} \epsilon ^3-32 M^{10} r^{16} \epsilon ^3+5 M^9
r^{17} (\epsilon -24) \epsilon ^2 \nonumber
\\
& +168 M^8 r^{18} \epsilon ^2-78 M^7 r^{19} \epsilon ^2+12 M^6
r^{20} (\epsilon -8) \epsilon +132 M^5 r^{21} \epsilon -60 M^4
r^{22} \epsilon \nonumber
\\
& +q^6 r^{11} \Big(M^3 \epsilon +r^3\Big)^2 \Big(7 M^3 \epsilon +4
r^3\Big)-3 M^3 r^{23} (8-3 \epsilon )+32 M^2 r^{24} \nonumber
\\
& -q^4 r^{12} \Big(M^3 \epsilon +r^3\Big)^2 \Big(40 M^4 \epsilon -19
M^3 r \epsilon +22 M r^3-10 r^4\Big) \nonumber
\\
& +q^2 r^{13} (2 M-r) \Big(M^3 \epsilon +r^3\Big)^2 \Big(38 M^4
\epsilon -17 M^3 r \epsilon +20 M r^3-8 r^4\Big)-14 M r^{25}\Big],
\end{align}
\begin{align}
P_{3}&=r^{13} \Big[8 a^2 \Big(q^2-M r\Big) \Big(M^3 \epsilon
+r^3\Big)^2 \Big(5 M^3 \epsilon +2 r^3\Big)+r^3 \Big(q^2 \Big(7 M^3
\epsilon +4 r^3\Big) \nonumber 
\\
& +r \Big(-12 M^4 \epsilon +5 M^3 r \epsilon -6 M r^3+2
r^4\Big)\Big)^2\Big],
\end{align}
\begin{align}
P_{4}&=\Big[a^2 r^{10} \Big(q^2-M r\Big)^2\Big(M^3 \epsilon
+r^3\Big)^6 \Big(a^2 \Big(M^3 \epsilon +r^3\Big)+r^3 \Big(-2 M
r+q^2+r^2\Big)\Big)^2 \Big(9 a^2 M^6 \epsilon ^2 \nonumber
\\
& +16 M^4 r^4 \epsilon -2 q^2 \Big(5 M^3 r^3 \epsilon +2 r^6\Big)-6
M^3 r^5 \epsilon +4 M r^7\Big)\Big]^{\frac{1}{2}}.
 \end{align}
The angular momentum is
\begin{align}
 \frac{L}{m}&=\pm\frac{1}{\sqrt{P_3} \beta}
\Bigg[ \Big(\mp2 a^2 r^6 (q^6+q^2 r^2M
(8M-5r)+rq^4(2r-5M)
 \nonumber
\\
& +M^2r^3(3r-4M))(r^3+\epsilon M^3)^4 +a^4r^3(q^2-Mr)(r^3+\epsilon
M^3)^4(2Mr^4-4M^4r\epsilon
 \nonumber
\\
& +q^2(-2r^3+M^3\epsilon))+P_{4}\Big)\sqrt{P_1+P_2}\Bigg].
\end{align}
where \begin{align}
 \beta&=a  r^3(M r-q^2)(M^3 \epsilon +r^3)^4
[2 r^3 \left(r (r-2 M)+q^2\right)^2+a^2(q^2(2 r^3-M^3 \epsilon )
\nonumber
\\
& +M r (4 M^3 \epsilon
 -3 M^2 r \epsilon -2
 r^3))].
 \end{align}
Here the upper sign indicates that the particle corotates with the black
hole while the lower sign refers to the opposite situation.

 In the case $\epsilon=0$,
the above expressions for the energy and angular momentum reduce to
the ones obtained for the Kerr-Newman metric \cite{dp}
\begin{align}
 &\frac{E}{m}=\frac{r^2-2Mr+q^2\pm a(Mr-q^2)^{1/2}}{r \sqrt{(r^2-3Mr+2q^2\pm 2a(Mr-q^2)^{1/2})}},
\\
& \frac{L}{m}=\pm \frac{(Mr-q^2)^{1/2}(r^2+a^2\mp
2a(Mr-q^2)^{1/2})\mp a q^2}{r \sqrt{(r^2-3Mr+2q^2\pm
2a(Mr-q^2)^{1/2})}}.
\end{align}
Taking $q=0=\epsilon$ gives the energy and angular momentum for the
Kerr metric \cite{bar}
\begin{align}
 &\frac{E}{m}=\frac{r^{3/2}-2Mr^{1/2}\pm aM^{1/2}}{r^{3/4}\sqrt{r^{3/2}-3Mr^{1/2}\pm 2aM^{1/2}}},
\\
& \frac{L}{m}=\pm \frac{M^{1/2}(r^2\mp
2aM^{1/2}r^{1/2}+a^2)}{r^{3/4}\sqrt{r^{3/2}-3Mr^{1/2}\pm2aM^{1/2}}}.
\end{align}
The ISCO is determined by numerically solving the equation
\begin{equation}
\frac{dE}{dr}=0.
\end{equation}
In Figure 4, radial dependence of ISCO is shown for some particular values of
the parameters.
\begin{figure}[!ht]
\centering
\includegraphics[scale=0.8]{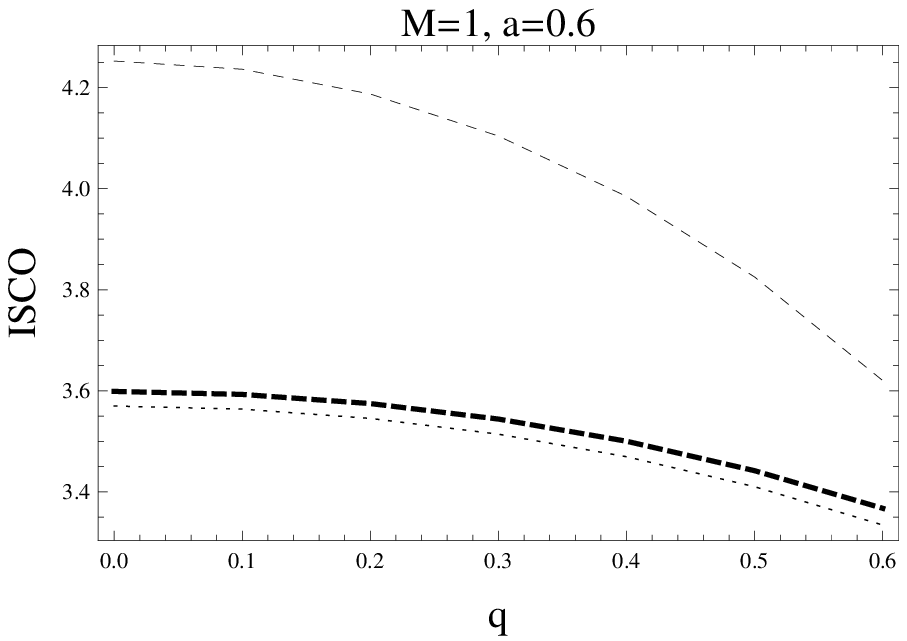}
\includegraphics[scale=0.8]{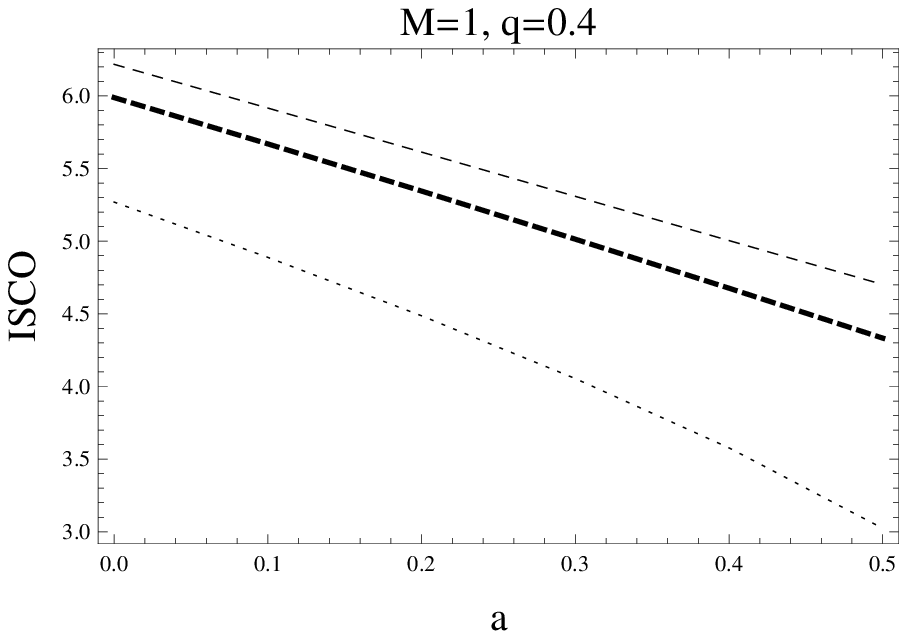}
\caption{The Inner most stable circular orbits for some values of
$\epsilon, a$ and $q$. Mass $M$ has been set to 1. On the left
$a=0.6$ with varying values of charge $q$. Starting from the upper
curve, the values of $\epsilon$ are $-1,1,2$ respectively. On the
right $q=0.4$ with varying values of the rotation parameter.
Starting from the upper curve, the values of $\epsilon$ are $-2,-1,2$
respectively.}
\end{figure}


The circualr photon orbits are located where
\begin{equation}
\frac{E}{m}\rightarrow \infty, \quad \frac{L}{m}\rightarrow \infty.
\end{equation}
In Figure 5, radial profile of the circular photon orbit is shown.
\begin{figure}[!ht]
\centering
\includegraphics[scale=0.8]{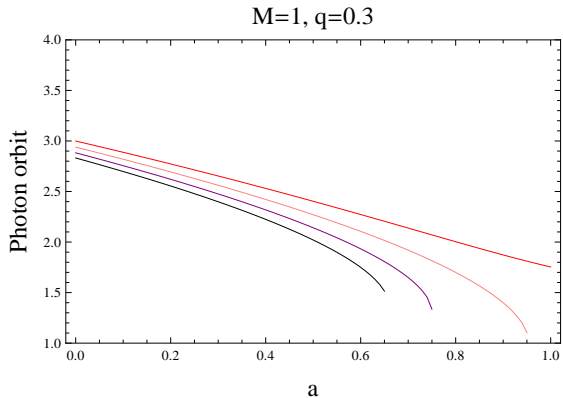}
\caption{The circular photon orbits for some values of $\epsilon$
and $a$. Mass $M$ has been set to 1 and $q$ is taken as $q=0.3$. Red
plot is for $\epsilon=-1$, pink for $\epsilon=0$, purple for
$\epsilon=1$, and black for $\epsilon=2$.}
\end{figure}
As in the ISCO case, increasing the value of the deformation
parameter decreases the radius of the photon orbits.

\section{Summary and conclusion}\label{SC}
Many modified theories of gravity have been
developed that lead to spacetimes that are different from Kerr's in that
they have parameters other than mass and spin. If one sets all the
deviations back to zero, such spacetimes reduce to Kerr.

One such spacetime has been formulated by Johannsen and Psaltis \cite{jp}. This metric takes
the deformed Schwarzschild spacetime as the starting point and applies Newman-Janis algorithm to obtain
a rotating spacetime. In this paper we have extended the approach to obtain the charged analogue of the Johannsen-Psaltis metric.
This has been done by taking the Reissner-Nordstr\"om spacetime as the seed metric and applying  Newman-Janis algorithm to it. This
leads to a spacetime which is stationary, axisymmetric and asymptotically flat. It gives the Johannsen-Psaltis metric on
setting charge equal to zero. Setting the deviation parameter $\epsilon=0$ yields the Kerr-Newman metric.
The electromagnetic 4-potential contains the deviation parameter $\epsilon$ in its radial component, thus inducing $\theta$ dependence in it.
 Thus $A_r$ component cannot be gauged away so we have a nonzero $A_r$  making the 4-potential $A_\mu$ different from the case of Kerr-Newman. Further analysis of $A_r$  revels that it depends on the mass $M$ through the
deviation parameter $\epsilon$ which we can consider as a coupling
between the gravitational and the electromagnetic fields.

The event horizon of the charged Johannsen-Psaltis spacetime has been studied.
Here horizon is taken as a general function \cite{sys} of $\theta$, $H(\theta)$, so that
horizon is $r=H(\theta)$. The equation which determines  $H(\theta)$ is given by (\ref{34}). Next,
the Killing horizon is discussed. The graphs are drawn for different values of
parameters $a,q,M,\epsilon$. The graphs show that for $\epsilon<\epsilon^{max-pole}$, the inner and outer horizons have spherical topology
while for  $\epsilon\approx\epsilon^{max-pole}$ Killing horizons intersect at the equatorial plane. In case
of $\epsilon>\epsilon^{max-pole}$, disjoint Killing horizons appear above and below the
equatorial plane. For negative $\epsilon$, the inner and outer Killing horizons are in the shape of the
spherical surface for $a^2+q^2<M$ while the shape changes to a toroidal surface if $a^2+q^2>M$. From the
determinant of the charged Johannsen-Psaltis metric we find that it is independent of the
Lorentz violating regions. The plots of the closed time-like curve and the Killing horizons show that
it is surrounded by the Killing horizon. Therefore, the charged Johannsen-Psaltis metric does not contain closed time-like curves outside of the central body.

Considering the charged Johannsen-Psaltis metric as a perturbation of the Kerr-Newman metric upto the first
order in $\epsilon$, we find that its Kretschmann scalar diverges at $r=0$, therefore it
represents a black hole for all values of $\epsilon$.

The analysis of the ISCO and circular photon orbits shows the dependence on $\epsilon$. Both show decreasing behavior with the increase of
$\epsilon$.

 \acknowledgments
RR is supported by the Indigenous Fellowship of the Higher Education Commission (HEC) of Pakistan.
The work was also supported by the HEC research grants  20-2087 and 6151, under its NRPU programme.

\end{document}